\documentstyle[12pt]{article}
\makeatletter
\begin{document}
\topmargin 0pt
\oddsidemargin -3.5mm
\headheight 0pt
\topskip 0mm \addtolength{\baselineskip}{0.20\baselineskip}
\begin{flushright}
 Alberta-Thy-09-99 \\
 hep-th/9910158
\end{flushright}
\vspace{0.5cm}
\begin{center}
  {\large \bf Comments on `` Black Hole Entropy from Conformal Field
  Theory in Any Dimension '' }
\end{center}
\vspace{0.5cm}
\begin{center}
by 
\vspace{0.3cm} \\ 
Mu-In Park$^a$\footnote{Electronic address: mipark@physics3.sogang.ac.kr} 
and Jeongwon Ho$^b$
\\ 
{\it $^a$ Department of Physics, Yonsei University, Seoul 120-749, Korea } \\ 
{\it  $^b$ Theoretical Physics Institute, Department of Physics, University of 
Alberta, Edmonton, Canada T6G 2J1} \\
\end{center}
\vspace{0.5cm}                              
\begin{center}
 {\bf ABSTRACT} 
 \end{center}
In a recent letter, Carlip proposed a generalization of the
Brown-Henneaux-Strominger construction to any dimension. We present 
two criticisms about his formulation. \\

\vspace{6cm}
%\vspace{0.5cm} 
\begin{flushleft}
% PACS Nos: 11.10.Ef, 11.10.Lm, 11.15.-q, 11.30.-j \\
September 1999 
\end{flushleft}

\newpage

Recently Carlip \cite{Car:99} proposed a generalization of the
Brown-Henneaux-Strominger construction to any dimension based on the
algebra of deformations at a black hole horizon. By restricting to
the horizon of a black hole, he has shown that the algebra contains a 
Virasoro subalgebra independent of the spatial dimensions with the appropriate
boundary conditions. With the aid of Cardy formula he derived the
Bekenstein-Hawking entropy within the conformal field theory. In this 
Comment, we point out that (a) his analysis has a
serious error which invalidates one of its principal claims and (b)
independent of that error there is one more serious defect.
 
The error (a) occurs in $\delta _{\xi_2} J [\hat{ \xi _1}] $ (10) of
Ref. [1] 
which essentially claims that it should be antisymmetric under $1
\leftrightarrow 2$ interchange according to his identification of
$\delta_{\xi_2} J[\hat{\xi}_1] =\{ J[\hat{\xi}_2], J[\hat{\xi}_1] \}^*
$ ( $\{, \}^*$ denotes the Dirac 
  bracket). But this is not correct identification because the
  antisymmetry can not be attained with his choice of $J[\hat{\xi}]$
  (9) and hence the Virasoro algebra, whose existence is one of its
  principal claims, can not be obtained in his
  form. Let us explain our conclusion in detail. We first note that the last two terms of $\delta _{\xi_2} J [\hat{ \xi _1}] $ have no manifest
  antisymmetricity for indices. In his paper
\cite{Car:99}, he ``assumed'' $\delta _{\xi _2} \hat{\xi ^t
  _1}=\hat{\xi}^r_2 \partial _r \hat{\xi }^t _1 $ to get the
antisymmetricity ``by hand''. But, this assumption is not acceptable
because one can show that $\delta _{\xi _2} \hat{\xi 
  ^t _1}=0$ which is essentially because $\delta_{\xi_2}
\hat{\xi}^t_1=(\delta_{\xi_2} N ) \xi^t_1 =0 $. Hence, the $\delta
_{\xi_2} J [\hat{ \xi _1}] $ is not antisymmetric under $1
\leftrightarrow 2$ because of the last two terms although it is antisymmetric
for the first two terms in (13). One can \underline{not} obtain the Virasoro
algebra with this preliminary algebra.

About the origin of this failure, we found that it is a direct
consequence of the {\it non-differentiability} of $L=H+J$ contrast to his claim. 
$\delta J$ does not cancel the boundary terms in $\delta
H$ exactly, rather it introduces several additional boundary terms in
$\delta L$: Most of these terms have no contribution on the horizon
because of his boundary condition $N^2 |_{r_+} =0$ but two of them can
not be neglected and these two contributions make $L$
non-differentiable. Let us explain first why the differentiability of
$L$ is connected to the antisymmetrization of $\delta_{\xi_2} J
[\hat{\xi}_1]$. When $L$ is the correct diffeomorphism ({\it Diff}) generator ,i.e., 
$\{ L[\eta], \phi \}=\delta_\eta \phi $ for any dynamical variables
  $\phi$, the {\it Diff} transformation of $L$ can be expressed as 
  $\delta_{\eta} L[ \xi]= \{ L[\eta], L[\xi] \}$. When the constraint 
  $H\approx 0$ is imposed this will reduce to : 
$\delta_\eta J[\xi]=\{ J[\eta], J[\xi] \}^*$ 
by picking out the boundary parts of $\delta_{\eta} L[ \xi]$ according to the
  definition of the Dirac bracket.
According to this identification, $\delta_\eta J[\xi]$ is
antisymmetric under $\eta \leftrightarrow \xi $ explicitly as $\delta_{\eta} L[ \xi]$ is. 
Now, let us consider the (general) variation of $L$ to investigate its
differentiability: 
\begin{eqnarray*}
\delta {L}
&=&  \frac{1}{8 \pi G} \int _{r_+} d^{n-2} x
\sqrt{\sigma} \left[ \frac{1}{2} \sigma^{\alpha \beta} n^r \hat{\xi
    }^t D_r \delta \sigma _{\alpha \beta} -n_r \hat{\xi}^r \delta K +
  \cdots \right] + \mbox{bulk} \nonumber \\
          &=& - \frac{1}{8 \pi G} \int d^{n-1} x
          \delta (r-r_+) \sqrt{\sigma}\left
          [ \frac{\sigma^{\alpha \beta}}{2}  \partial_r (n^r
          \hat{\xi}^t) \delta g_{\alpha \beta}+ 
          \frac{n_r \hat{\xi}^r
          }{(2-n) f \sqrt{\sigma} } g_{rr} \delta \pi^{rr} + \cdots
          \right] 
          +\mbox{bulk},
\end{eqnarray*}
where the contributions of `\dots' terms to the {\it Diff} transformation 
$\delta_{\eta} \phi$ are order of $O (N^2)$ or $O(N)$ which vanish on the
horizon. From these two
non-vanishing boundary contributions, we get the corresponding $anomalous$
boundary contributions in the following {\it Diff} transformations:
\begin{eqnarray*}
\delta_{\xi} g_{rr}(x)&=&\{ L[\hat{\xi}], g_{rr} (x) \}  \\
                &=&\frac{1}{8 \pi G}\delta (r-r_+)
\frac{n_r \hat{\xi}^r }{(2-n) f \sqrt{\sigma} } g_{rr} +\mbox{bulk} , \\
\delta_{\xi} \pi^{\alpha \beta} (x) &=& \{L[\hat{\xi}], \pi^{\alpha
                \beta}(x) \} \nonumber \\ 
                &=&- \frac{1}{8 \pi G} \delta (r-r_+)
                \frac{\sqrt{\sigma}}{2} \sigma^{\alpha \beta}
                (n^r
          \hat{\xi}^t) 
          +\mbox{bulk},
\end{eqnarray*}
whose boundary contributions behave as $\delta(r-r_+) \times O(1)$
which can not be neglected.
Here, the bulk part will represent the usual {\it Diff} transformations
in the bulk. These wrong {\it Diff} transformations implies that Carlip's original choice of $J
[\hat{\xi}]$ is not correct one to make $L[\hat{\xi}]$ be the {\it Diff} generator  for the system with the horizon boundary and to give the claimed Virasoro
algebra. %We don't have alternative solution yet.
 
Criticism (b) consists of the observation that even upon ignoring
point (a), his scenario does not work for the non-rotating black hole even though it works well for the rotating case. Recently 
a resolution has been suggested in a different context of 2D gravity 
by Cadoni and Mignemi \cite{Cad:99} who have introduced the 
time-integrated generator instead of (usual) fixed-time generator. 
But we do not think that this can be resolved even in the way of Cadoni
and Mignemi although the time integration of (21) for
the non-rotating case gives the similar result as the rotating case with 
different central charge
$c=3 A\beta/(\pi G T)$ and different $L_0 =AT/(8 \pi G \beta)$ \cite{Park:99}. 
A strange fact is that the resulting entropy 
$
\mbox{log}\rho(L_0)=2 \times \frac{A}{4 G} 
$
has a wrong factor of 2 which remind us a similar wrong factor in their 
computation of 2D-gravity entropy \cite{Cad:99}. More serious problem of 
this method is that the time integration of the left-hand sides of (21) can 
\underline{not} be written as
$
\{\bar{J} [\hat{\xi}_m],\bar{J} [\hat{\xi}_n] \}^*
$
for time-integrated generator $\bar{J} =\frac{1}{\tau} \int^{\tau}_0 dt
J(t)$, which is essential for the interpretation of (21) as a
Virasoro algebra.

MIP would like to thank Prof. Steven Carlip for stimulus discussions along this work and Prof. Jae Hyung Yee, Drs. Gung-won Kang and 
Hyuk-jae Lee for several helpful discussions. MIP was supported in part by a 
postdoctoral grant from
the Natural Science Research Institute, Yonsei University in the year
1999 and in part by the Korea Research Foundation under 98-015-D00061. JH was 
supported in part by Korea Science and Engineering Foundation and in part by 
National Science and Engineering Research Council of Canada.

\end{document}